\documentclass[11pt]{article}
\usepackage{amsfonts}

\def\P{{\mathbb P}}
\def\S{{\mathcal S}}
\def\X{{\mathcal X}}

\usepackage{amssymb}
\usepackage{amsmath}

\title{Predicting upcoming actions by observation: some facts, models and challenges}

\author{C.~D.~Vargas$^{1}$, M~.L.~Rangel$^{1}$ and
  A.~Galves$^{2}$\\
  $^{1}$ Instituto de Biof\'\i sica Carlos Chagas Filho, Universidade Federal do Rio de Janeiro\\
 $^{2}$Instituto de Matem\'atica e Estat\'\i stica, Universidade de S\~ao Paulo}

\date{May 21, 2014}

\begin{document}

\maketitle

\begin{abstract}

  Predicting another person's upcoming action to build an appropriate
  response is a regular occurrence in the domain of motor control. In
  this review we discuss conceptual and experimental approaches aiming
  at the neural basis of predicting and learning to predict upcoming
  movements by their observation.
\end{abstract}

\noindent{\it Key words}: Action anticipation, bayesian approach, statistical
model selection, motor prediction

\noindent {\it Short title}: Predicting actions by observation

\noindent {\it Corresponding author}: C.~D.~Vargas
 Instituto de Biof\'\i sica Carlos Chagas Filho, Universidade Federal do Rio de Janeiro, CCS-Bloco G, Ilha do Fund\~ao 21941-902, Rio de Janeiro, RJ, Brazil,  {\tt cdvargas@biof.ufrj.br}

\section{Introduction}

How does the brain predict an incoming movement? This requires an
efficient coding of the movement features, including the context in
which it occurs.  As matter of fact, predicting movements in a
variable environment is a fundamental aspect of skilled motor behavior
(cf., for instance, \cite{Kawato1987}, and \cite{MiallW96}).

Prediction is essentially a statistical inference task.  This is in
part due to the noise associated to stimuli detection and
processing as well as to the intrinsic stochasticity of the brain
functioning. This is also due to the fact that predicting means
choosing the next state of the upcoming movement given the knowledge
of the past steps and of the inferred context. This choice is
guided by an evaluation of the probabilities of the action's possible outcomes.

Statistical inference requires a probabilistic model. This means a set
of possible outcomes and a probability measure assigning a number
between 0 and 1 to every event that can be expressed in this set of
outcomes. In the case of motor prediction the brain must operate with
a probabilistic model evolving in time, incorporating the past
experience of the agent.  This time evolution should allow to model
the capacity of plastic change and adaptation found in motor systems,
which is maintained throughout life by learning and
learning-to-predict mechanisms.

This leads to a set of basic questions. First of all, how does the
brain build the probabilistic models used to perform the statistical
inference tasks required by motor prediction, and, more generally, to
produce actions?  How do these models evolve in time so as to
incorporate former experiences and corresponding inference results?
How should the model represent the network structure involved in the
inference task? Discussing these questions from a neuromathematical
point of view is the goal of the present text.

In Section \ref{predicting_session} we present a set of experiments
aiming at the neural basis of estimating upcoming actions performed by
others. In Section \ref{bayes_session} we briefly review how the
bayesian perspective takes this experimental framework into account.
In Section \ref{futur} we show that the bayesian approach is just an
exemple of a more general framework, namely the statistical model
selection approach. Finally, in a conclusion section, we briefly
discuss new perspectives and challenges in modeling action prediction.

\section{Predicting actions through observation: some neurophysiological 
evidence}\label{predicting_session}

Voluntary movements consist in transforming an action goal into a
movement appropriate to a given context (review in
\cite{Wolpert01}). In humans, this task is achieved by a widespread
network of brain areas recruited both before and during action
execution. A subset of these areas is also active whenever one
observes, simulates or imagines an action, even without any explicit
motor output (review in \cite{Jeannerod}). Within this framework,
anticipating other agents' motor behavior from the observation of
their ongoing actions could lead to the recruitment of neural circuits
similar to those enrolled in motor planning and execution. An
extensive review of this field is outside the scope of this
paper. Thus, we will herein briefly visit a set of experiments
addressing such conjecture.

Kilner {\sl et al.} (2004) investigated whether the readiness
potential, traditionally described as an electrophysiological marker
of motor preparation (\cite{Deecke}, \cite{Shibasaki}), could also be
detected whenever an observer expected an upcoming action to occur in
a visual display \cite{KilnerVargas}. The experiment consisted in
randomly presenting videos depicting hands either in resting position
or while grasping a glass. Before the beginning of the experiment,
subjects were informed that each time the glass was green the actor
would grab it, whereas whenever it was red, no action would be depicted
on the display. Results showed a readiness potential in the green
condition, that is, when both the nature and onset time of the
upcoming action was predictable \cite{KilnerVargas}. In contrast, no
readiness potential was evoked in the red condition. These results
suggested that the mere knowledge of a coming action automatically
activates the motor system. The authors concluded that, besides being
a marker of motor preparation, the readiness potential might also be
regarded as a neural correlate of motor prediction
\cite{KilnerVargas}.

In a very similar paradigm, Fontana {\sl et al.} (2012) investigated
the contribution of two key regions in anticipating upcoming actions
performed by others \cite{Fontana}. The authors examined whether a
readiness potential was generated in chronic stroke patients with
focal lesions either in the parietal or the premotor cortex when they
expected to observe an upcoming movement in a visual display. They
found that the readiness potential was preserved in the patients with
premotor lesions but not in those with parietal lesions. These results
suggested that the parietal cortex integrity is important in the
capacity of estimating the occurrence of upcoming actions performed by
others. As the parietal cortex is thought to create forward models of
intended movements (\cite{Desmurget}, \cite {Andersen},
\cite{BlakemoreSirigu03}), the failure found in parietal patients in
anticipating the onset of an observed action was interpreted as due to
an impaired action prediction mechanism.

In another line of evidence, Aglioti {\sl et al.} (2008)
applied transcranial magnetic stimulation in the primary motor cortex
to investigate the dynamics of action anticipation and its underlying
neural correlates in professional basketball players \cite{Aglioti}. The participants
were exposed to videos presenting basketball players performing
correct (IN) and incorrect (OUT) free shots. The results showed an
increased motor evoked potential response in hand muscles of elite
athletes (but not in novices nor in basketball expert observers) for
the OUT shots. This indicates that high levels of motor expertise are
linked to the fine-tuning of specific anticipatory resonance
mechanisms. Furthermore, these results strongly suggested that these
anticipatory mechanisms are dependent on a previous motoric experience
\cite{Aglioti}. Taken together, this set of results indicates that the
ability to anticipate upcoming movements performed by other agents (or
else, anticipating their outcomes) draws on brain areas enrolled in
motor control (\cite{KilnerVargas}, \cite{Aglioti}), certainly
involves prior learning (\cite{Aglioti}, \cite{Grafton},
\cite{Petroni} ) and seems to depend on the parietal cortex's integrity
\cite{Fontana}.

Now the question is how does the brain process actions being performed
by other agents to anticipate/predict upcoming movements.  The
bayesian framework has been suggested as a possible approach to model
the way the brain chooses a most likely outcome, using both its past
experience and the new online information provided by external stimuli
(cf., for instance, \cite{Kording}, \cite{Knill}, \cite{Doya} and
\cite{Orban}).  In the next session this framework will be briefly
presented.

\section{The bayesian approach}\label{bayes_session}

The brain operates in the presence of incomplete evidence. It must
assign patterns to observed actions, either to react in an appropriate
way, or just to give meaningful interpretations to observed
scenes. The incomplete evidence is provided by sensory
information. Using this incomplete and noisy information, the brain
must make predictions about the {\sl state of the world}.

To express this situation mathematically we must define a
probabilistic model. The first component of the probabilistic model is
the set $\Omega$ of possible outcomes. The set $\Omega$ is
traditionally called the {\sl sample space}. In the present case
$\Omega$ is a set of ordered pairs. The first element of the pair is
the hidden variable, say $s$, expressing a {\sl state of the world}
that it is not directly observed. The second element of the pair, say
$x$, is the available observed information about this hidden variable
provided the sensory stimuli. Therefore, if we call $\S$ the set of
hidden variables and $\X$ the set of observed variables, we have that
the sample space $\Omega$ is the cartesian product of $\S$ and $\X$
\[
\Omega=\S \times\X\, .
\]

The second ingredient of a probabilistic model is a probability
measure $\P$ associated to the sample space. In the bayesian framework
this probability measure is defined as a mixture of measures. More
precisely, $\P$ is defined as a weighted combination of conditional
probabilities on $\X$,  with weights defined through a
distribution on the set of hidden variables $\S$. If we assume that
the sets of hidden and observed variables are countable sets, then this can 
be expressed as follows. For any element $x \in \X$, we have
\begin{equation}\label{mixture}
\P (x)=\sum_{s \in \S} \P(x\, |\, s) \P(s).
\end{equation}

The above formula was written assuming that the sets $\S$ and $\X$ are
finite or at most countable. However, usually motor control data
assumes real or even function values. In this case, and in general,
whenever the sets $\S$ and $\X$ are not countable, the sum in formula
\ref{mixture} must be replaced by an integral and $\P(x\,|\, s)$,
interpreted as a density function.

The distribution on $\S$ which defines the weights in formula
\ref{mixture} is called the {\sl a priori} distribution. The {\sl a
  priori} distribution contains the previous information about the
statistical distribution of the states of the world available before
the experience producing the observed variables takes place. The
conditional probability $\P(x\,|\, s)$ tells how likely it is to get
the observable value $x$, given that the hidden state of the world is
$s$.

To ilustrate these notions consider the case of a duel, in which a
swordsman faces an armed opponent. In this case, predicting the motion
of the opponent's sword is really a matter of life or death. The
observable variable is the one given by the observation of the motion
of the opponent's sword. This observation is noisy and contains a
certain amount of imprecision. Previous knowledge of fencing
techniques tells the swordsman how to interpret the hidden intentions
of his opponent. Even better if he has previous knowledge of his
opponent's style. All this previous knowledge is embodied in the {\sl
  a priori} distribution. The prediciton of the hidden goal of the
opponent is made by evaluating the possible goals of the opponent,
taking into account the observable movement of the opponent and
evaluating these possibilities using as weights the {\sl a priori}
knowledge of his opponent's style.

Formally this is done using formula (\ref{mixture}) and the definition
of conditional probability

\begin{equation}\label{bayes}
\P(s\, |\, x)=\frac{\P(s)}{\P(x)}\P(x\,|\, s).
\end{equation}

Equation (\ref{bayes}) is the famous Bayes formula. It shows that the
best prediciton about the hidden state of the world, given the
observable evidence $x$,  is the state $s$ which maximizes
$\P(s)\P(x\,|\, s)$. 

This shows how the available evidence is evaluated using the previous
knowledge contained in the {\sl a priori} distribution. Given the
empirical evidence $x$, the distribution $\P(\cdot\, |\, x)$ on the
set of hidden states of the world $\S$ is called the {\sl a
  posteriori} distribution. It encapsulates our new knowledge about
the probability distribution of the hidden states of the world, taking
into account the new empirical evidence provided by $x$.

At this point it is interesting to make a comparison between the
bayesian approach to statistics and the {\sl classical} frequentist
point of view. In the {\sl classical} frequentist approach to
statistics, the parameters of the unknown probality measure are
estimated using the Law of Large Numbers, and the Central Limit
Theorem is used to obtain confidence intervals or regions for these
estimations. The Law of Large Numbers and the Central Limit Theorem
are asymptotical results. They hold when the sample size diverges to
infinity. In other terms, the quality of the estimation increases with
the size of the sample.

An estimation based on an asymptotical result is not what the
swordsman really needs. He cannot wait until he has a large sample of
movements of his opponent to estimate what is the goal
purchased by his opponent sword in his first and maybe final attack. He
must defend himself immediately. That is what makes the bayesian
approach much more attractive to him.

In the bayesian approach the estimation is done immediately. The
swordsman uses his previous knowledge of his opponent's style and
fencing techniques as weights to evaluate what is his opponent's most
likely goal, given the empirical evidence contained in the visible
data $x$.

Both the swordsman and the brain must make decisions in real time,
without waiting until a large sample of evidence is available. This
does not mean that the accumulated past experience is not taken into
account. Actually, every time new empirical evidence arrives, we
update our knowledge about the probability distribution on the set
$\S$ of the states of the world. The previous {\sl a priori}
distribution is replaced by the {\sl a posteriori} distribution
obtained using the visible variable provided experimentally. This {\sl
  a posteriori} distribution becomes the new {\sl a priori}
distribution, and so on.

To exemplify, let us now rephrase Kilner {\sl et al} (2004) and
Fontana {\sl et al} (2012) in bayesian terms (\cite{KilnerVargas},
\cite{Fontana}). In this case, the set of {\sl states of the world}
$\S$ has two elements: {\sl action} $a$ and {\sl no action} $n$. The
set of observable variables $\X$ also has two elements: {\sl green
  condition} $g$ and {\sl red condition} $r$.

The {\sl a priori} distribution implicitly considered in Kilner {\sl
  et al} (2004) and Fontana {\sl et al} (2012) assigns equal
probabilities $1/2$ and $1/2$ to each state of the world. In bayesian
terminology, an {\sl a priori} distribution which assigns equal
probabilities to all the states of the world is called {\sl non
  informative}.

Finally, the conditional probabilities $\P(\cdot\,|\, a)$ and
$\P(\cdot\,|\, n)$ are degenerated, in the sense that each one of
them assigns probabilites $1$ or $0$ to each one of the observable
variables, namely
\[ 
\P(g\,|\, a)=1 \, \, \mbox{and} \, \, \P(r\,|\, a)=0
\] 
and 
\[ 
\P(g\,|\, n)=0 \, \, \mbox{and} \, \, \P(r\,|\, n)=1\, .
\]

In Kilner {\sl et al} (2004) and Fontana {\sl et
  al} (2012), the conditional probabilities are
implicitly defined as follows. Given that the state of world is {\sl
  action}, the probability of having the {\sl green condition} is $1$
and the probability of having the {\sl red condition} is $0$. On the
other hand, given that the state of world is {\sl no action}, the
probability of having the {\sl red condition} is $1$ and probability
of having the {\sl green condition} is $0$.

Using now Bayes formula (\ref{bayes}), we now compute the {\sl a
  posteriori} probability and obviously they are also
degenerated. Namely
\[ 
\P(a \,|\, g)=1 \, \, \, \, \, \mbox{and}\, \,\P(n\,|\, r)=0
\] 
and 
\[ 
\P(a \,|\, r)=0 \, \,\mbox{and}\, \,\P(n\,|\, n)=1\, .
\] 
In other terms, in this experimental setup, the {\sl action} is
predicted with probability $1$, whenever the observed variable was the
{\sl green condition}, and the state of {\sl no action} is predicted
with probability $1$, given that the observed variable was the {\sl
  red condition}.

This results in a situation where, after several repetitions per
condition, a measurable parameter (the readiness potential) was
extracted from the brain signal in the green condition but not in the
red condition. This was taken as an evidence that the readiness potential would
be a marker of predictive coding \cite{KilnerVargas}.  In the case of
parietal patients this predictive mechanism would be supposedly
dysfunctional. This is coherent with the experimental result which
shows no marker of motor preparation to incoming movements performed
by other agents (i.e, the green condition) for this group of patients \cite{Fontana}.

There is a practical reason to work with events which have 
probability either $0$ or $1$. The reason is the difficulty to identify
evoked potentials trial by trial. As a matter of fact, the experiments
presented above use a large number of perfectly synchronized trials to
extract a readiness potential from the otherwise noisy EEG signals. In
these cases the analysis strategy consists in making an average of the
signals obtained in the repetitions so as to eliminate the
fluctuations of the EEG signal. Therefore, with this approach, only
events which occur with probability $1$ or $0$ will be extracted from
the experimental data. To get evidence
of the occurrence of events with a probability strictly between $0$
and $1$, one should be able to identify the evoked response in a
trial by trial basis. This requires new signal processing procedures
as the ones proposed, for instance, by \cite{quiroga2000} or by
\cite{AbootalebiMK09}.

Now a more general question is: how to infer from this or any other
electrophysiological data the mechanisms by which the brain builds
such estimates. Any attempt to model the brain's inference activity
must be able to describe the way this inference machine evolves in
time and incorporates previous experiences. This issue will be
discussed in the next section.

The bayesian predictive solution described above is an example of a
more general framework. Namely, given a sample of observable
variables, how to assign a model for the source producing this
data. In statistical terms this is called {\sl statistical model
  selection}. This suggests a new paradigm not only for motor learning
and prediction, but more generally to describe the way the brain
encodes and processes information.

\section{Modeling the brain activity using statistical model
  selection}\label{futur}

In Aglioti {\sl et al.} (2008), the corticospinal excitability of hand
muscles was modulated in elite basketball players whenever an OUT shot
was about to occur \cite{Aglioti}. This was taken as an evidence that
the motor system was {\sl estimating} the result of the shot
beforehand. Such effect was absent both in trained observers and in
novice players, suggesting that the learning and retrieval of such
motor representations is crucial to bring forth the brain activity
corresponding to the next upcoming movement in the context of
observation.

This experiment can be rephrased as a statistical model selection
procedure performed throughout time. The
observer encodes in a suitable way each successive step of the action
being performed by the player in the movie. Each step is encoded by a
symbol sumarizing the main features of the observed action. This
translates the movie into a symbolic chain. This chain is
intrinsically random, in other terms, the movie is encoded as a
realization of a stochastic chain. Acting as a statistician the brain
uses the information provided by the successive steps of the movie to
identify in an adaptive way the probabilistic structure of the
stochastic chain producing the sample. And then it uses the transition
probabilities identifyed in this way to predict the successive steps
of the action.

Let us state this in a more formal way. Denote by $A$ the set of all
symbols used to encode each step of the observed action. To simplify, we assume that the set $A$ is finite. For each$n=1,2,\ldots$, let $X_n$ denote the symbol belonging to the set $A$
which was used to encode the $n^{th}$ step of the action. 

The sequence of symbols
\[
X_1, X_2, X_3,\ldots,
\]
produced in this way can be interpreted as a realization of a {\sl
  stochastic chain}, which is a mathematical term to denote a discrete
time evolution affected by chance. The question is which stochastic
chain should be used to fit in an economic way the observed symbolic
chain. This is precisely the goal of a statistical model selection
procedure.

Model selection involves the choice of a class of candidate models and
the choice of a procedure to select a member of this class, given the
data. Stochastic chains with memory of variable length are good
candidates to represent in an economic way random stationary sources
producing sequences of symbols.

Chains with memory of variable length appear in Rissanen's paper
called {\sl A universal data compression system}
(\cite{rissanen}). His idea was to model a string of symbols as a
realization of a stochastic chain where the length of the memory
needed to predict the next symbol is not fixed, but it is a 
function of the string of the past symbols.

It turns out that in many data sets consisting of the strings of symbols,
 the length of the relevant
portion of the past is not fixed, on the contrary it depends on the
past.  Rissanen's ingenious idea was to construct a stochastic model
that generalizes this notion of relevant portion of the past to any
kind of symbolic strings. This obviously includes symbolic chains
obtained by encoding the successive steps of the action being
performed by a basketball player and any other kind of action.

In (\cite{rissanen}) Rissanen called {\it context} the relevant part
of the past. The law of the stochastic chain is defined by the set of all
contexts and an associated family of transition probabilities which gives
the probability of the next symbol, given a context. The set of
contexts expresses in a precise and economic way the structural
dependencies present in the data. 

Besides the choice of a class of candidate models, model selection
also requires the choice of a procedure to select a member of the
class of candidate models. For the class of chains with memory of
variable lenght this issue has been addressed by an increasing number
of papers, starting with \cite{rissanen} who introduced the so-called
{\sl Algorithm Context}. For a recent result on
this area and a more extensive bibliography we refer the reader to
\cite{GGGGL}.

The Algorithm Context works in an adaptive way. It estimates at each
step the length of the context associated to the string of symbols
observed until that time step, as well as the associated transition
probability. In other terms, the Algorithm Contexts checks at each
step how much of the past information present in the observed chain is
relevant to predict the next step. It gives as an output the shortest
suffix of the past symbols obtained in this way as a candidate
context. This procedure is done in real time and when the sample is
big enough, the algorithm identifies in a precise way each one of
the contexts characterizing the chain producing the sample.

It is tempting to conjecture that the adaptive procedure performed by
the Algorithm Context mimics in some sense the statistical procedure
performed by the brain when it selects in real time a model for the
action being performed by another agent. And if this picture is correct,
then the model for the observed action, in other terms, the neural
code assigned to the action, can be described as a set of contexts 
and a family of associated transition probabilities.

\section{Challenges and perspectives}

Statistical model selection in suitable classes of stochastic chains
should be a major notion in neuroscience modeling. Research in motor
control and prediction would particularly benefit from modeling
actions as realisations of stochastic chains. Within this framework,
the prediction of the next steps would be performed throught time, in
an adaptive way, by assigning more and more effective models based on
a statistical procedure strongly reminiscent of the Algorithm
Context. Such approach might likely allow inferring about the
mechanisms by which the brain predicts upcoming actions.

Implementing this point of view requires the introduction of new
classes of stochastic systems describing the time evolution of neural
networks in simple and economic way. These models must be simple
enough to be analysed in a mathematically rigourous way. At the same
time, these models must display some of the important qualitative
features of these networks. This combination of simple models with
complex and realistic behavior is essential to make progress in the
understanding of the brain activity.  An example of a simple
mathematical model of this type was recently introduced in Galves and
L\"ocherbach (2013)(\cite{galves_loecherbach_2013}). The development
of these new classes of stochastic systems is a challenge for
mathematicians.

In parallel, the development of this new statistical model selection
paradigm requires the invention of new procedures allowing to
challenge this paradigm from an experimental point of view. What kind
of experimental evidence would reject the assumption that the brain
learns and interpret the world by performing a model selection
procedure? To the best of our knowledge, neuroscience is still
 not able to
describe how the brain could possibly perform statistical
analyses. Making progress in this direction is a major challenge for
neuroscientists.

\section*{Acknowledgments}

This work is part of CAPES/Nuffic project 038/12, USP project {\sl
  Mathematics, computation, language and the brain} and was done as an
activity of FAPESP's {\sl Research, Inovation and Dissemination Center
  for Neuromathematics-NeuroMat} (FAPESP grant 2011/51350-6). CDV and
AG are partially supported by CNPq fellowships  (303247/2011-8 and
309501/2011-3 respectively) and CNPq grants (480108/2012-9 and
478537/2012-3 respectively). CDV is also supported by FAPERJ (grant
E26/110.526/2012). CDV thanks Numec-USP and AG thanks INDC for
hospitality.
\bibliography{biblio}
\end{document}